\documentclass[amstex,aps,preprint,showpacs,nofootinbib]{revtex4}
\usepackage{graphicx}
\usepackage{amsmath}
\usepackage{bm}
\usepackage{amsfonts}
\usepackage{amssymb}

\begin{document}

\title{Bell's theorem as a signature of nonlocality:\\a classical counterexample.}
\author{A.\ Matzkin}
\affiliation{Laboratoire de Spectrom\'{e}trie Physique (CNRS Unit\'{e} 5588),
Universit\'{e} Joseph-Fourier Grenoble-1, BP 87, 38402 Saint-Martin
d'H\`{e}res, France}

\begin{abstract}
For a system composed of two particles Bell's theorem asserts that averages of
physical quantities determined from local variables must conform to a family
of inequalities. In this work we show that a classical model containing a
local probabilistic interaction in the measurement process can lead to a
violation of the Bell inequalities. We first introduce two-particle
phase-space distributions in classical mechanics constructed to be the analogs
of quantum mechanical angular momentum eigenstates. These distributions are
then employed in four schemes characterized by different types of detectors
measuring the angular momenta. When the model includes an interaction between
the detector and the measured particle leading to ensemble dependencies, the
relevant Bell inequalities are violated if total angular momentum is required
to be conserved. The violation is explained by identifying assumptions made in
the derivation of Bell's theorem that are not fulfilled by the model. These
assumptions will be argued to be too restrictive to see in the violation of
the Bell inequalities a faithful signature of nonlocality.

\end{abstract}
 \maketitle

\section{Introduction}

Bell's theorem was originally introduced \cite{bell1964,bb} to examine
quantitatively the consequences of postulating hidden variable distributions
on the incompleteness of quantum mechanics put forward by Einstein, Podolsky
and Rosen \cite{EPR} (EPR). In particular, the hidden variables were supposed
to locally and causally complete quantum mechanics by making sense of the
'reality' of physical quantities described by non-commuting operators relative
to two spatially separated particles in an entangled state. Bell showed that a
correlation function obtained from averages over the hidden variables of these
physical quantities must satisfy certain inequalities (the Bell inequalities),
and that these inequalities are violated by quantum mechanical averages. Given
that experiments have confirmed with increasing precision the correctness of
the quantum formalism, it is generally stated that the violation of the Bell
inequalities contradicts locality. The strong version of such statements
asserts that quantum mechanics itself is non-local \cite{stapp}.\ This vocable
is quite popular (in particular among non-specialists as well as in quantum
information papers) but there is a general agreement among most specialists
that this strong assertion is unsubstantiated \cite{percival,mermin,unruh}.
Instead, the received view is the weak version following which Bell's theorem
asserts the incompatibility of local hidden variables with quantum mechanics.
Nevertheless it can objected, in principle \cite{fine99,accardi} or through
abstract models \cite{orlov,christian}, whether the assumptions made in order
to derive Bell's theorem are necessary in order to enforce locality, or
whether they only rule out a certain manner of ascribing local variables to
the measurement outcomes.

In this work we will show that statistical distributions in classical
mechanics can violate Bell-type inequalities. Moreover the statistical
distributions we will employ are not exotic objects but the \emph{classical
analogues} of the quantum-mechanical coupled angular momenta eigenstates, so
that our model is essentially the classical version of the paradigmatic
2-particles singlet state. The violation of the inequalities can of course be
achieved only provided the model falls outside the assumptions necessary in
order to prove Bell's theorem. This role will be played by a probabilistic
interaction that is assumed to take place between the measured particle and
the detector, combined with the requirement that the total angular momentum be
conserved. Although this interaction is local, it nevertheless spoils the
derivation of Bell's theorem, because it introduces an ensemble dependency of
the outcomes: the resulting averages then involve correlations given by
conditional probabilities between ensembles rather than between the individual
phase-space positions. As a consequence, the different expectation values
employed in Bell's inequalities cannot be derived jointly, as required in the
derivation of the theorem.

The paper is organized as follows.\ We will start by introducing the
classical phase-space distributions (Sec.\ II), first for a single
particle, then for 2 particles with total zero angular momentum. We
will explain why these distributions are the classical analogues of
the quantum angular momentum eigenstates. In Sec.\ III we will
investigate three different examples of Bell-type models.\ Each of
the examples will be characterized by the same phase-space
distribution but by differing detection schemes.\ In the first case,
the projections of the angular momentum of each of the particles
along arbitrary axes are directly measured by the detectors, leading
to a straightforward application of Bell's theorem (which will be
briefly derived). In the second example the detectors yield discrete
outcomes, depending on the values of the angular momenta; this
example, which also abides by Bell's theorem, will allow us to
introduce conditional probabilities to account for the correlated
angular momenta. The third example will illustrate the same
situation with stochastic variables (the angular momenta specify
probabilities of obtaining an outcome). In Sec.\ IV, we will
introduce an example falling outside the class of Bell-type models.\
This example will also involve discrete measurement outcomes, but
the presence of an interaction leading to ensemble dependencies will
be introduced. We will see that ensemble dependencies lead to
non-commutative measurements for a single particle, and to the
violation of the Bell inequalities for initially correlated
two-particle systems. In Sec. V we will discuss these results,
insisting on the role played by the existence of joint distributions
and on the relationship between locality and conservation laws. A
short summary and our conclusion are given in Sec.\ VI.

\section{Classical distributions analogues of angular momenta eigenstates}

\subsection{One-particle angular momentum distributions}

A quantum mechanical angular momentum eigenstate $\left|  j_{0}m\right\rangle
$ is characterized by a well-defined value $\sqrt{j_{0}(j_{0}+1)}$ of the
modulus of the angular momentum $\mathbf{J}$ and of its projection $J_{z}$ (of
value $m)$ along a given axis $z$. In configuration space the spherical
harmonic $\left|  \left\langle \theta,\phi\right|  \left.  j_{0}m\right\rangle
\right|  ^{2}$ gives the probability distribution corresponding to a fixed
value of $J$ and $J_{z}$ as $\theta$ and $\phi$ (the polar and azimuthal
angles) span the unit sphere. The classical statistical distributions can be
considered either in phase-space, defined by $\Omega=\{\theta,\phi,p_{\theta
},p_{\phi}\}$ where $p_{\theta}$ and $p_{\phi}$ are the conjugate canonical
momenta, or in configuration space. Let us assume the modulus $J$ of the
angular momentum is fixed.\ Let $\rho_{z}(\Omega)$ be the distribution in
phase-space given by%
\begin{equation}
\rho_{z_{0}}(\theta,\phi,p_{\theta},p_{\phi})=N\delta(J_{_{z}}(\Omega
)-J_{z_{0}})\delta(J^{2}(\Omega)-J_{0}^{2}). \label{5}%
\end{equation}
$\rho_{z_{0}}$ defines a distribution in which every particle has an angular
momentum with the same magnitude, namely $J_{0}$, and the same projection on
the $z$ axis $J_{z_{0}}$, without any additional constraint. Hence
$\rho_{z_{0}}$ can be considered as a classical analog of the quantum
mechanical density matrix $\left|  j_{0}m\right\rangle \left\langle
j_{0}m\right|  $. Eq. (\ref{5}) can be integrated over the conjugate momenta
to yield the \emph{configuration space} distribution
\begin{equation}
\rho(\theta,\phi)=N\left[  \sin(\theta)\sqrt{J_{0}^{2}-J_{z_{0}}^{2}/\sin
^{2}(\theta)}\right]  ^{-1} \label{7}%
\end{equation}
where we have used the defining relations $J_{_{z}}(\Omega)=p_{\phi}$ and
$J^{2}(\Omega)=p_{\theta}^{2}+p_{\phi}^{2}/\sin^{2}\theta$ to perform the
integration. Further integrating over $\theta$ and $\phi$ and requiring the
phase-space integration of $\rho$ to be unity allows to set the normalization
constant $N=J_{0}/2\pi^{2}$.\

$\rho(\theta,\phi)$ gives the statistical distribution of the particles in
configuration space.\ Its standard graphical representation (parameterization
on the unit sphere) is shown in Fig.\ 1(a) along with the quantum mechanical
orbital momentum eigenstate (a spherical harmonic taken for the same values of
$j$ and $m$) in Fig. 1(b). The similarity of both figures is a statement of
the quantum-classical correspondence in the semiclassical regime, since
$\sqrt{\rho(\theta,\phi)}$ is approximately the amplitude of the configuration
space quantum mechanical eigenstate for high quantum numbers. Note that rather
than working with the particle distributions in configuration space, it will
also be convenient to visualize the distribution of the angular momentum in
\emph{physical} space corresponding to a given particle distribution (see Fig.
1(c)); $\theta$ and $\phi$ will then denote the position of $\mathbf{J}$ on
the the angular momentum sphere.

Let us take a second axis $a$ making an angle $\theta_{a}$ relative to the $z
$ axis (in this paper we will take all the axes to lie in the $zy$ plane). We
can define a distribution by fixing the projection $J_{a}$ of the angular
momentum on $a$ to be constant, $\rho_{a_{0}}=\delta(J_{a}-J_{a_{0}}%
)\delta(J-J_{0}^{2})$. In configuration space, this distribution $\ $may be
shown to be obtained by rotating the distribution of Eq. (\ref{7}) by the
angle $\theta_{a}$ towards the $a$ axis.\ We will be interested below in
determining the average projection $J_{a}$ on the $a$ axis for a distribution
of the type (\ref{7}) corresponding to a well defined value of $J_{z}$. Using
$J_{a}=J_{z}\cos\theta_{a}+J_{y}\sin\theta_{a}$, $J_{z}=p_{\phi}$ and
\begin{equation}
\left\langle J_{y}\right\rangle _{J_{z_{0}}}=\int J\sin\theta\sin\phi
\sin\theta_{a}\delta(p_{\phi}-J_{z_{0}})d\Omega=0
\end{equation}
by rotational invariance, we obtain%
\begin{equation}
\left\langle J_{a}\right\rangle _{J_{z_{0}}}=\int p_{\phi}\cos\theta_{a}%
\delta(p_{\phi}-J_{z_{0}})d\Omega=J_{z_{0}}\cos\theta_{a}. \label{6}%
\end{equation}

\begin{figure}[tb]
\includegraphics[height=2.in,width=4.7in]{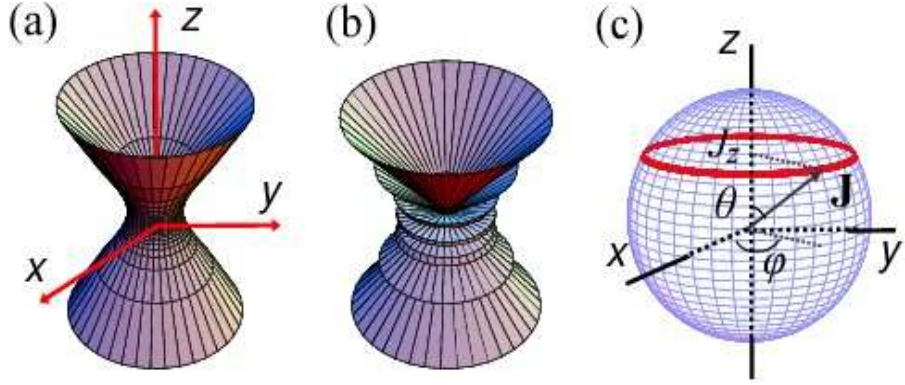}  \caption{Normalized
angular distribution for a single particle in configuration space. (a)
\emph{Classical} distribution $\rho(\theta,\phi)$ of Eq. (\ref{7}). (b)
Corresponding \emph{quantum} distribution (spherical harmonic $|Y_{JM}%
(\theta,\phi)|^{2}$ with $J/\eta=40$, $\eta=\hbar$, and $M/J=5/8$ as in (a)).
(c) Distribution of the angular momentum on the sphere for a distribution of
the type $\rho(\theta,\phi)$, invariant around the $z$ axis with a fixed value
of $J_{z}$. }%
\label{f1}%
\end{figure}Note that a given $\mathbf{J}$ can belong jointly to several
distributions $\rho_{a_{0}}$ and $\rho_{b_{0}}$ ($a$ and $b$ being
different directions).\ But if we require that any distribution must
correspond to a well-defined value of the angular momentum
projection along a given axis, then distributions such as
$\rho_{a_{0}}$ and $\rho_{b_{0}}$ become mutually exclusive. The
ring-like distribution of the angular momentum on the angular
momentum sphere represented in Fig.\ 1(c) (corresponding to the
configuration space distribution shown in Fig. 1(a)) can be
generalized to cover the entire hemisphere centered on the $z$ axis
(see Fig.\ 3(a)).\ Then properties such as $J_{1a}$ and $J_{1b}$
being of the same sign on such hemispheres become mutually exclusive
properties.

\subsection{Two-particle angular momentum distributions}

The situation we will consider below, by analogy with the well-known EPR-Bohm
pairs in quantum mechanics, is that of the fragmentation of an initial
particle with a total angular momentum $\mathbf{J}_{T}=0$ into 2 particles
carrying angular momenta $\mathbf{J}_{1}$ and $\mathbf{J}_{2}$. Conservation
of the total angular momentum imposes $J_{1}=J_{2}\equiv J$ and%
\begin{equation}
\mathbf{J}_{1}+\mathbf{J}_{2}=0. \label{e1}%
\end{equation}
Eq. (\ref{e1}) implies a \emph{correlation}, imposed initially at the source,
between the angular momenta of the 2 particles and of their projections along
any axis $a$: the knowledge of the value of $J_{1a}$ allows to infer the value
of $J_{2a}$, $J_{2a}=-J_{1a}$. Without further constraints (or additional
knowledge), the classical distribution in the 2-particle phase space is given
by%
\begin{equation}
\rho(\Omega_{1},\Omega_{2})=N\delta(\mathbf{J}_{1}+\mathbf{J}_{2})\delta
(J_{1}^{2}-J^{2}), \label{9}%
\end{equation}
where $N$ is again a normalization constant. The corresponding distributions
of the angular momenta in physical space -- easier to visualize than $\rho$ --
is uniform on the sphere, with $\mathbf{J}_{1}$ and $\mathbf{J}_{2}$ pointing
in opposite directions (see Fig. 2(a)), reflecting the isotropic character of
the fragmentation as well as the correlation (\ref{e1}).

\section{Bell-type models}

\subsection{Setting}

The Bell inequalities are obtained by computing average values of
measurement outcomes performed independently on each of the 2
particles. Three examples are studied below, all involving the
initial fragmentation of a particle with zero angular momentum
(Sec.\ II.B). In the first example, we assume that the measurements
give directly the value of the projection of the angular momentum of
each particle along an arbitrarily chosen axis. In the second
example we introduce detectors having a threshold, resulting in
discrete measurement outcomes depending solely on the position of
the particles' angular momenta. The third example is a repetition of
the second but with stochastic variables. Bell's theorem, which is
derived in Sec. III.B, is verified in all these cases. To alight the
notation, we will choose units such that $J=1$.

\subsection{Bell's theorem}

\subsubsection{Example 1: direct measurement of the classical angular momenta}

Two particles with initial total angular momentum $\mathbf{J}_{T}=0$ flow
apart. Let $a$ and $b$ be two axes in the $zy$ plane. The projection of
particle 1's angular momentum along the $a$ axis, $J_{1a}$ and that of
particle 2 along $b$, $J_{2b}$ are measured. The average of the joint
measurement outcomes on the 2 particles is directly given by the values of
$J_{1a}$ and $J_{2b}$ and the probability distribution given by Eq.
(\ref{9}).\ All these quantities depend on the phase-space position of the
particles, i.e. on the position of the angular momenta on the sphere (see
Fig.\ 2(a)). The average is computed from%
\begin{equation}
\left\langle J_{1a}J_{2b}\right\rangle =\int J_{1a}(\Omega_{1})J_{2b}%
(\Omega_{2})\rho(\Omega_{1},\Omega_{2})d\Omega_{1}d\Omega_{2}. \label{3}%
\end{equation}
Given the rotational symmetry, $z$ is chosen along $a$, hence $J_{1a}%
=p_{\phi_{1}}$ and
\begin{equation}
J_{2b}=p_{\phi_{2}}\cos\left(  \theta_{b}-\theta_{a}\right)  +\left\{
J\sin\theta_{2}\sin\phi_{2}\sin\left(  \theta_{b}-\theta_{a}\right)  \right\}
.
\end{equation}
One first integrates over $\phi_{2}$ (the term between $\{..\}$ vanishes) then
over $p_{\phi_{2}}$ (yielding $p_{\phi_{2}}=-p_{\phi_{1}}$ because of the
correlation $\delta(J_{1a}+J_{2a})$).\ The last non-trivial integration is
over $p_{\phi_{1}}$,%
\begin{equation}
\left\langle J_{1a}J_{2b}\right\rangle =\int_{-1}^{1}dp_{\phi_{1}}-p_{\phi
_{1}}^{2}\cos\left(  \theta_{b}-\theta_{a}\right)  \left[  2\pi N\int
d\tilde{\Omega}\right]  \label{e5}%
\end{equation}
where $d\tilde{\Omega}$ represents the variables remaining after the
integration of the delta functions. Since $\rho$ is normalized, we have%
\begin{equation}
\int_{-1}^{1}dp_{\phi_{1}}2\pi N\int d\tilde{\Omega}=1. \label{e7}%
\end{equation}
Integrating Eq. (\ref{e7}) over $p_{\phi_{1}}$ allows to obtain the value
between the $\left[  ..\right]  $ in Eq. (\ref{e5}) thereby avoiding the
explicit calculation of the normalization constant. The result for the
expectation is
\begin{equation}
E(a,b)\equiv\left\langle J_{1a}J_{2b}\right\rangle =-\frac{1}{3}\cos\left(
\theta_{b}-\theta_{a}\right)  .
\end{equation}

\subsubsection{Derivation of the Bell inequality}

The correlation function $C(a,b,a^{\prime},b^{\prime})$ involved in Bell's
inequality is given by%
\begin{equation}
C(a,b,a^{\prime},b^{\prime})=\left(  \left|  E(a,b)-E(a,b^{\prime})\right|
+\left|  E(a^{\prime},b)+E(a^{\prime},b^{\prime})\right|  \right)  /V_{\max
}^{2}\label{23}%
\end{equation}
where $a^{\prime}$ and $b^{\prime}$ are arbitrary axes in the $xy$ plane and
$V_{\max}$ is the maximal absolute value that can be obtained in a measurement
outcome. Let us denote by $V_{1a}(\Omega_{1})$, $V_{2b}(\Omega_{2})$ etc. the
detected values along the relevant axes, with the 2-particle average being%
\begin{equation}
E(a,b)=\int V_{1a}(\Omega_{1})V_{2b}(\Omega_{2})\rho(\Omega_{1},\Omega
_{2})d\Omega_{1}d\Omega_{2}.\label{26}%
\end{equation}
The Bell inequality%
\begin{equation}
C(a,b,a^{\prime},b^{\prime})\leq2\label{25}%
\end{equation}
puts a bound on the value of the correlation function. It is obtained
\cite{bell-bert} by forming the difference
\begin{equation}
E(a,b)-E(a,b^{\prime})=\int V_{1a}(\Omega_{1})\left[  V_{2b}(\Omega
_{2})-V_{2b^{\prime}}(\Omega_{2})\right]  \rho(\Omega_{1,}\Omega_{2}%
)d\Omega_{1}d\Omega_{2}\label{e10}%
\end{equation}
where $V_{1a}$ has been factored. Likewise,%
\begin{equation}
E(a^{\prime},b)+E(a^{\prime},b^{\prime})=\int V_{1a^{\prime}}\left[
V_{2b}+V_{2b^{\prime}}\right]  \rho d\Omega_{1}d\Omega_{2}.\label{e11}%
\end{equation}
We now use $\left|  V_{2\beta}\right|  \leq V_{\max}$ ($\beta=b,b^{\prime}$)
to derive
\begin{equation}
\left|  V_{2b}-V_{2b^{\prime}}\right|  +\left|  V_{2b}+V_{2b^{\prime}}\right|
\leq2V_{\max}.\label{e12}%
\end{equation}
Take the absolute values and use $\left|  V_{1\alpha}\right|  \leq V_{\max}$
($\alpha=a,a^{\prime}$) in each of the Eqs. (\ref{e10}) and (\ref{e11}) to
obtain two inequalities.\ Adding these inequalities and using (\ref{e12})
leads to the Bell inequality%
\begin{equation}
\left|  E(a,b)-E(a,b^{\prime})\right|  +\left|  E(a^{\prime},b)+E(a^{\prime
},b^{\prime})\right|  \leq2V_{\max}^{2}.\label{e13}%
\end{equation}
In the present example, $V_{\max}=1$, and $C(a,b,a^{\prime},b^{\prime})$ is
bounded by $2\sqrt{2}/3$, so that the Bell inequality (\ref{25}) is verified.

As a corollary, note that the factorization made in Eqs. (\ref{e10}%
)-(\ref{e11}) is equivalent \cite{fine82} to the existence of joint
distributions of the form%
\begin{equation}
\mathcal{F}_{aba^{\prime}b^{\prime}}=\int V_{1a}(\Omega_{1})V_{2b}(\Omega
_{2})V_{1a^{\prime}}(\Omega_{1})V_{2b^{\prime}}(\Omega_{2})\rho(\Omega
_{1},\Omega_{2})d\Omega_{1}d\Omega_{2}. \label{e60}%
\end{equation}
Indeed, Bell's inequality can be proved \cite{clauser-shimony} by adding and
substracting $\mathcal{F}_{aba^{\prime}b^{\prime}}$ from Eq. (\ref{e10}) and
then factorizing $V_{1a}V_{2b}$ and $V_{1a}V_{2b^{\prime}}$ respectively. The
term $\mathcal{F}_{aba^{\prime}b^{\prime}}$ is the average obtained when 4
measurements are made -- 2 outcomes are obtained for each particle (particle's
1 $V$ property is measured along the axes $a$ and $a^{\prime}$ whereas
particle 2 is measured along the axes $b$ and $b^{\prime}$). The
factorization, or equivalently the existence of $\mathcal{F}_{aba^{\prime
}b^{\prime}}$, is an important assumption in the derivation of the inequalities.

\subsubsection{Derivation in the stochastic case and joint distributions}

In the stochastic case, a given phase-space position $(\Omega_{1},\Omega_{2})$
does not determine a unique valued outcome $(V_{1a}(\Omega_{1}),V_{2b}%
(\Omega_{2}))$ as above (corresponding to what is usually termed
''deterministic case'') but determines instead well-defined probabilities
$p(V_{1a},V_{2b},\Omega_{1},\Omega_{2})$ of obtaining $(V_{1a},V_{2b})$. The
counterpart of the factorization made in Eq. (\ref{e10}) lies in the
factorization of the probabilities,%
\begin{equation}
p(V_{1a},V_{2b},\Omega_{1},\Omega_{2})=p(V_{1a},\Omega_{1})p(V_{2b},\Omega
_{2}),\label{w1}%
\end{equation}
where $p(V_{1a},\Omega_{1})$ is the single particle elementary probability
such that
\begin{equation}
P(V_{1a})=\int p(V_{1a},\Omega_{1})\rho(\Omega_{1})d\Omega_{1}.\label{w6}%
\end{equation}
The expectation value (\ref{26}) is then replaced by%
\begin{equation}
E(a,b)=\int\bar{V}_{1a}(\Omega_{1})\bar{V}_{2b}(\Omega_{2})\rho(\Omega
_{1},\Omega_{2})d\Omega_{1}d\Omega_{2}\label{w2}%
\end{equation}
with%
\begin{align}
\bar{V}_{1a}(\Omega_{1}) &  =\sum V_{1a}p(V_{1a},\Omega_{1})\label{w3}\\
\bar{V}_{2b}(\Omega_{2}) &  =\sum V_{2b}p(V_{2b},\Omega_{2}).\label{w4}%
\end{align}
The derivation leading to Eq. (\ref{e13}) proceeds as above by replacing the
value of the outcomes by their respective averages $\bar{V}_{1a}$ and $\bar
{V}_{2b}$. The factorization (\ref{w1}) allows to obtain a joint probability
for an arbitrary number of events from the elementary probabilities
$p(V,\Omega);$ the counterpart to Eq. (\ref{e60}) is%
\begin{equation}
P_{aba^{\prime}b^{\prime}}=\int p(V_{1a},\Omega_{1})p(V_{2b},\Omega
_{2})p(V_{1a^{\prime}},\Omega_{1})p(V_{2b^{\prime}},\Omega_{2})\rho(\Omega
_{1},\Omega_{2})d\Omega_{1}d\Omega_{2}.\label{w5}%
\end{equation}
Note that the existence of a joint probability $P_{aba^{\prime}b^{\prime}}$
(that appears here as a consequence of the factorization (\ref{w1})) leads
immediately to the inequality (\ref{e13}) irrespective of \emph{any other
assumption} concerning the dependence of the outcomes or probabilities on
supplementary variables (here the phase-space positions, the
'hidden-variables' in quantum mechanics). Indeed, using expressions of the
type%
\begin{equation}
E(a,b)=\sum_{V_{1a},V_{2b}}V_{1a}V_{2b}\sum_{V_{1a^{\prime}},V_{2b^{\prime}}%
}P_{aba^{\prime}b^{\prime}}%
\end{equation}
for the average values, we have%
\begin{equation}
\left|  E(a,b)-E(a,b^{\prime})\right|  \leq\sum P_{aba^{\prime}b^{\prime}%
}\left|  V_{1a}\left(  V_{2b}-V_{2b^{\prime}}\right)  \right|
\end{equation}
and an analog inequality for $\left|  E(a^{\prime},b)+E(a^{\prime},b^{\prime
})\right|  $.\ Adding both inequalities yields
\begin{align}
\left|  E(a,b)-E(a,b^{\prime})\right|  + &  \left|  E(a^{\prime}%
,b)+E(a^{\prime},b^{\prime})\right|  \leq\nonumber\\
&  \sum P_{aba^{\prime}b^{\prime}}\left(  \left|  V_{1a}\left(  V_{2b}%
-V_{2b^{\prime}}\right)  \right|  +\left|  V_{1a^{\prime}}\left(
V_{2b}+V_{2b^{\prime}}\right)  \right|  \right)  \leq2V_{\max}^{2}%
\text{,}\label{i1}%
\end{align}
since the expression between $(...)$ is bounded by $2V_{\max}^{2}$ and the
joint probability sums to 1.

\subsection{Discrete outcomes\label{exd}}

In this second example, we take over the setup of the first example except for
the measurement outcomes: we now assume that a given detector placed on an
axis can only give 2 values, depending on the sign of the angular momentum's
projection.\ The outcomes are given by
\begin{equation}
D_{1a}(\Omega_{1})=\left\{
\begin{tabular}
[c]{l}%
$\frac{1}{2}\text{ if }J_{1a}>0$\\
$-\frac{1}{2}\text{ if }J_{1a}<0$%
\end{tabular}
\ \ \right.  \text{ \ \ }D_{2b}(\Omega_{2})=\left\{
\begin{tabular}
[c]{l}%
$\frac{1}{2}\text{ if }J_{2b}>0$\\
$-\frac{1}{2}\text{ if }J_{2b}<0$%
\end{tabular}
\ \right.  \label{e15}%
\end{equation}
and depend only on the positions $\mathbf{J}_{1}$ and $\mathbf{J}_{2}$ of the
angular momentum (hence on the phase-space position of the measured
particles). The average value
\begin{equation}
\left\langle D_{1a}D_{2b}\right\rangle =\int D_{1a}(\Omega_{1})D_{2b}%
(\Omega_{2})\rho(\Omega_{1},\Omega_{2})d\Omega_{1}d\Omega_{2} \label{e17}%
\end{equation}
takes the form%
\begin{equation}
\left\langle D_{1a}D_{2b}\right\rangle =\sum_{k,k^{\prime}=-1/2}%
^{1/2}kk^{\prime}\int_{\mathcal{D}(k,k^{\prime})}\rho d\Omega_{1}d\Omega_{2}
\label{e20}%
\end{equation}
where $\mathcal{D}(k,k^{\prime})$ represents the domain of integration on
which the joint conditions \textrm{sign}($J_{2a})=-\mathrm{sign}%
(J_{1a})=-\mathrm{sign}(k)$ \emph{and} \textrm{sign}($J_{2b})=\mathrm{sign}%
(k^{\prime}) $ hold (see Fig. 2(b)). The integral gives the probability
\begin{equation}
P_{kk^{\prime}}\equiv P(D_{1a}=k\cap D_{2b}=k^{\prime})=P(D_{1a}%
=k)P(D_{2b}=k^{\prime}|D_{1a}=k) \label{e21}%
\end{equation}
where $P(D_{2b}=k^{\prime}|D_{1a}=k)$ is the probability of obtaining
$D_{2b}=k^{\prime}$ conditioned on the knowledge that $D_{1a}=k$. The
conditional probability appears because of the initial correlation (\ref{e1})
-- the positions of the angular momenta are not independent. The conditional
probability can more easily be determined on the angular momentum sphere by
computing the area where $\mathrm{sign}(J_{2b})=\mathrm{sign}(k^{\prime})$
relative to the area of the hemisphere where $\mathrm{sign}(J_{2a}%
)=-\mathrm{sign}(k)$ (of area $2\pi$). This area is given by the intersection
of the two relevant hemispheres (see Fig. 2(b)), i.e. a spherical lune whose
area can be put under the form $2\pi k(k-k^{\prime})+4kk^{\prime}(\theta
_{b}-\theta_{a})$. Since $\rho$ is uniform on the sphere, we have
$P(D_{1a}=k)=1/2$ from where%
\begin{equation}
P_{kk^{\prime}}=k(k-k^{\prime})+\frac{2kk^{\prime}}{\pi}\left|  \theta
_{b}-\theta_{a}\right|  ,
\end{equation}
and the average $\left\langle D_{1a}D_{2b}\right\rangle $ becomes%
\begin{equation}
E(a,b)=-\frac{1}{4}+\frac{\left|  \theta_{b}-\theta_{a}\right|  }{2\pi}.
\end{equation}
The maximal detected value here is $V_{\max}=1/2$. The correlation function is
computed from Eq. (\ref{23}) and it may be verified that $C(a,b,a^{\prime
},b^{\prime})$ is bounded by $2$: Bell's inequality (\ref{25}) is again verified.

\subsection{Discrete outcomes: a stochastic model}

We now elaborate on the preceding example to give a model in line with the
stochastic version of Bell-type variables. A given position of the angular
momentum of a particle in phase-space does not specify the outcome $S$, as in
Eq. (\ref{e15}), but the probabilities $p(S_{1a}=k,\Omega_{1})$ of obtaining
the outcome $k$. For definiteness we will replace Eqs. (\ref{e15}) by%
\begin{align}
p(S_{1a} &  =\frac{1}{2},\Omega_{1})=\left\{
\begin{tabular}
[c]{l}%
$\frac{3}{4}\text{if }J_{1a}>0$\\
$\frac{1}{4}\text{ if }J_{1a}<0$%
\end{tabular}
\ \ \ \right.  \text{ \ \ }p(S_{2b}=\frac{1}{2},\Omega_{2})=\left\{
\begin{tabular}
[c]{l}%
$\frac{3}{4}\text{if }J_{2b}>0$\\
$\frac{1}{4}\text{ if }J_{2b}<0$%
\end{tabular}
\ \ \ \right.  \label{z1}\\
p(S_{1a} &  =-\frac{1}{2},\Omega_{1})=\left\{
\begin{tabular}
[c]{l}%
$\frac{1}{4}\text{if }J_{1a}>0$\\
$\frac{3}{4}\text{ if }J_{1a}<0$%
\end{tabular}
\ \ \ \right.  \text{ \ \ }p(S_{2b}=-\frac{1}{2},\Omega_{2})=\left\{
\begin{tabular}
[c]{l}%
$\frac{1}{4}\text{if }J_{2b}>0$\\
$\frac{3}{4}\text{ if }J_{2b}<0$%
\end{tabular}
\ \ \ \right.  .\label{z2}%
\end{align}
The expectation value involves first averaging, for each phase space position,
over the two possible outcomes, before averaging over the distribution $\rho$
of the angular momenta:%
\begin{equation}
\left\langle S_{1a}S_{2b}\right\rangle =\int\bar{S}_{1a}(\Omega_{1})\bar
{S}_{2b}(\Omega_{2})\rho(\Omega_{1},\Omega_{2})d\Omega_{1}d\Omega
_{2}\label{z5}%
\end{equation}
with (cf.\ Eqs. (\ref{w2})-(\ref{w4}))%
\begin{align}
\bar{S}_{1a}(\Omega_{1}) &  =\sum_{k}kp(S_{1a}=k,\Omega_{1})\\
\bar{S}_{2b}(\Omega_{2}) &  =\sum_{k^{\prime}}k^{\prime}p(S_{2b}=k^{\prime
},\Omega_{2}).
\end{align}
Taking into account the correlation at the source [Eqs. (\ref{e1})-(\ref{9})],
we proceed as in the preceding example, except that now each probability
$P_{kk^{\prime}}$ contains several contributions with a weight given by
$p(S_{1a}=k,\Omega_{1})p(S_{2b}=k^{\prime},\Omega_{2})$ that depends, through
Eqs. (\ref{z1})-(\ref{z2}), on the domains $\mathcal{D}(\pm1/2,\pm1/2)$ over
which \textrm{sign}($J_{1a})=\mp1$ and \textrm{sign}($J_{2b})=\pm1$.\ For
example for $k,k^{\prime}=\frac{1}{2}$, we have%
\begin{equation}
P_{\frac{1}{2}\frac{1}{2}}=\frac{9}{16}\int_{\mathcal{D}(\frac{1}{2}%
,\frac{1}{2})}\rho d\Omega_{1}d\Omega_{2}+\frac{3}{16}\int_{\mathcal{D}%
(\frac{1}{2},-\frac{1}{2})}\rho d\Omega_{1}d\Omega_{2}+\frac{1}{16}%
\int_{\mathcal{D}(-\frac{1}{2},\frac{1}{2})}\rho d\Omega_{1}d\Omega
_{2}+\frac{3}{16}\int_{\mathcal{D}(-\frac{1}{2},-\frac{1}{2})}\rho d\Omega
_{1}d\Omega_{2};
\end{equation}
now each integral represents a probability $P(\mathrm{sign}(J_{1a})=\mp
1\cap\mathrm{sign}(J_{2b})=\pm1)$. Comparing with Eqs. (\ref{e21}%
)-(\ref{e21}), we see that in the stochastic case, the probabilities
$P_{kk^{\prime}}$ depend as in the preceding example on the areas on the
angular momentum sphere occupied by the individual positions of each angular
momentum compatible with the outcomes (although in the stochastic case there
are many more such areas, each contributing with a given weight). Overall, Eq.
(\ref{z5}) yields%
\[
\left\langle S_{1a}S_{2b}\right\rangle =\sum_{k,k^{\prime}=-1/2}%
^{1/2}kk^{\prime}P_{kk^{\prime}}=\frac{1}{8}\left(  \frac{\theta_{b}%
-\theta_{a}}{\pi}-1\right)  .
\]
$C(a,a^{\prime},b,b^{\prime})$ is readily computed and is again, in line with
Bell's theorem, bounded by $2$.

\begin{figure}[tb]
\includegraphics[height=2.in,width=3.2in]{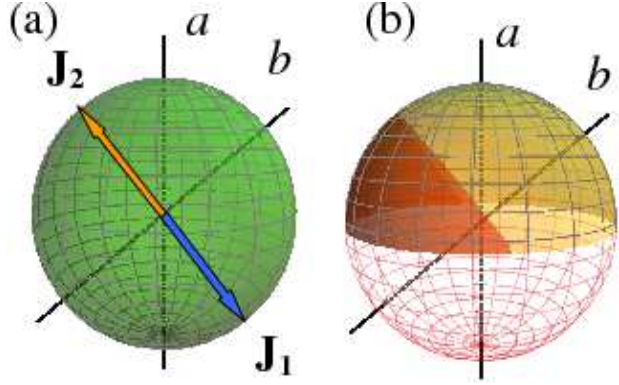}
%
%\label{f1}%
\caption{(a) Uniform distribution of $\mathbf{J}_{1}$ and $\mathbf{J}_{2}$ on
the unit sphere; the angular momenta are correlated via the conservation law
(\ref{e1}) and must thus point in opposite directions. In the first example
(Sec. III.B), the detectors measure $J_{1a}$ and the correlated $J_{2b}$ as
the angular momenta span the sphere. (b) Example 2 (Sec. III.C): Distribution
of $\mathbf{J}_{2}$ , when $D_{1a}=-1/2$ was obtained. A measurement of
$D_{2b}$ will yield $\pm1/2$ depending on the position of $\mathbf{J}_{2}$: if
$\mathbf{J}_{2}$ lies within the light shaded region (intersection of the two
positive hemispheres centered on $a$ and on $b$, denoted $\mathcal{D}%
(-1/2,1/2)$ in the text), $D_{2b}=1/2$ will be obtained, $-1/2$ when
$\mathbf{J}_{2}$ belongs to the dark-shaded region ($\mathcal{D}(-1/2,-1/2)$). }%
\end{figure}

\section{A detection model violating the inequalities}

The fourth example has similarities and differences with the models studied in
Secs. III.C and III.D.\ A given detector on an axis measures the angular
momentum's projection but only delivers the outcomes $\pm1/2$. However, the
outcomes depend on a probabilistic random interaction between the detected
particle and the detector. This interaction has a specific property (it
vanishes on average) that results in the introduction of an ensemble
dependency. We will see that this feature combined with the conservation of
the angular momentum between ensembles prevents the factorization that was
seen above to be necessary in order to derive Bell's theorem.

\subsection{Particle-detector interaction for a single particle\label{sp}}

\subsubsection{Basic properties}

Let $\rho_{1}(\Omega_{1})$ be the phase-space distribution for the single
particle 1 and $R_{1a}=\pm1/2$ denote the outcome obtained by placing a
detector on the $\ a$ axis. Let $P(R_{1a}=k,\rho_{1})$ be the probability of
obtaining the reading $k$ on the detector if the statistical distribution of
particle 1 (or equivalently, the distribution of $\mathbf{J}_{1}$) is known to
be $\rho_{1}$. We will impose the following constraint on the interaction: the
average $\left\langle J_{1a}\right\rangle _{\rho_{1}}$ over phase-space of the
measured value is the one obtained by averaging over the measurement outcomes.
This constraint takes the form%
\begin{equation}
\left\langle R_{1a}\right\rangle _{\rho_{1}}=\sum_{k=-1/2}^{1/2}%
kP(R_{1a}=k,\rho_{1})=\left\langle J_{1a}\right\rangle _{\rho_{1}}, \label{50}%
\end{equation}
meaning that whereas individual outcomes depend on the interaction,
on average the net effect of this interaction is zero. The models
leading to Eq. (\ref{50}) are not unique -- any model verifying Eq.
(\ref{50}) and obeying $\sum_{k}P(R_{1a}=k,\rho_{1})=1$ will do.
Depending on the specific model, Eq. (\ref{50}) will not be verified
for an arbitrarily chosen distribution $\rho_{1}$; only a class of
distributions can be consistent within a given model. In the present
model, we will assume as in the previous examples that $\rho_{1}$
can only be a uniform distribution occupying one (or both) of the
two hemispheres of the angular momentum sphere.

Let us examine for such distributions the consequences of Eq. (\ref{50}).
Assume that $\rho_{1}$ corresponds to a uniform distribution of $\mathbf{J}%
_{1}$ on the positive hemisphere centered on the $a$ axis, to be denoted
$\rho_{1a+}$. If a measurement is made along the $b$ axis, a direct
computation of $\left\langle J_{1b}\right\rangle _{\rho_{1a+}}$ gives%
\begin{equation}
\left\langle R_{1b}\right\rangle _{\rho_{1a+}}=\sum_{k}kP(R_{1b}=k,\rho
_{1a+})=\frac{1}{2}\cos\left(  \theta_{b}-\theta_{a}\right)  . \label{e25}%
\end{equation}
If one measures $R_{1a}$ the average (\ref{e25}) becomes $+1/2$, i.e. the only
positive detected outcome. Therefore, since the probabilities are positive, we
must have
\begin{align}
P(R_{1a}  &  =1/2,\rho_{1a+})=1\label{e27}\\
P(R_{1a}  &  =-1/2,\rho_{1a+})=0. \label{e29}%
\end{align}
Conversely if the distribution is $\rho_{1a-}$ (uniform on the lower
hemisphere) we obtain the opposite probabilities
\begin{align}
P(R_{1a}  &  =1/2,\rho_{1a-})=0\label{27}\\
P(R_{1a}  &  =-1/2,\rho_{1a-})=1 \label{29}%
\end{align}
and
\begin{equation}
\left\langle R_{1b}\right\rangle _{\rho_{1b-}}=\sum_{k}kP(J_{1b}=k,\rho
_{1a-})=-\frac{1}{2}\cos\left(  \theta_{b}-\theta_{a}\right)  . \label{e30}%
\end{equation}
Note that Eq. (\ref{e25}) along with the normalization of the
probabilities uniquely determines the value of the probabilities
\begin{equation}
P(R_{1b}=\pm\frac{1}{2},\rho_{1a+})=\frac{\cos\left(  \theta_{b}-\theta
_{a}\right)  \pm1}{2} \label{esp}%
\end{equation}
as well as the equality between the relative expectation value corresponding
to positive (resp. negative) outcomes $R_{1b}$ and the average of the angular
momentum projection over the regions where $J_{1b}$ is positive (resp.
negative), i.e.%
\begin{equation}
\pm\frac{1}{2}P(R_{1b}=\pm\frac{1}{2},\rho_{1a+})=\left\langle H(\pm
J_{1b})J_{1b}\right\rangle _{\rho_{1a+}}, \label{50b}%
\end{equation}
$H$ denoting the unit-step function.

The main property of this particle-detector interaction based model is that
the detected result does not depend on a phase-space point (or on a given
individual position of the particle's angular momentum on the sphere), be it
through a deterministic value ascription or through well-defined
probabilities. Indeed, if this were the case, then Eqs. (\ref{e27})-(\ref{29})
would imply that
\begin{equation}
R_{1a}(\Omega_{1})=1/2\Leftrightarrow J_{1a}>0\quad\text{and}\quad
R_{1a}(\Omega_{1})=-1/2\Leftrightarrow J_{1a}<0, \label{e33}%
\end{equation}
as in the example studied in Sec. III.D. But then assume $R_{1b}$ is measured
and the ensemble is known to be $\rho_{1a+}$ (uniform distribution on the
positive hemisphere centered on the $a$ axis). On the angular momentum sphere
$\rho_{1a+}$ can be seen as being composed of the intersections with
$\rho_{1b+}$ and $\rho_{1b-}$, $\rho_{1a+}=(\rho_{1a+}\cap\rho_{1b+})\cup
(\rho_{1a+}\cap\rho_{1b-})$. The respective integration domains are
$\mathcal{D}(-\frac{1}{2},\frac{1}{2})$ and $\mathcal{D}(-\frac{1}%
{2},-\frac{1}{2})$ (we use the notation introduced in Sec.\ III.C; see Fig.
2(b))). Hence%
\begin{equation}
\left\langle R_{1b}\right\rangle _{\rho_{1a+}}=\int R_{1b}(\Omega_{1}%
)\rho_{1a+}(\Omega_{1})d\Omega_{1}=\frac{1}{2}\int_{\mathcal{D}(-\frac{1}%
{2},\frac{1}{2})}\rho_{1b+}(\Omega_{1})d\Omega_{1}-\frac{1}{2}\int
_{\mathcal{D}(-\frac{1}{2},-\frac{1}{2})}\rho_{1b-}(\Omega_{1})d\Omega_{1},
\end{equation}
yielding $(1-2(\theta_{b}-\theta_{a})/\pi)/2$ in contradiction with the
constraint (\ref{50}) defining the model, $\left\langle J_{1b}\right\rangle
_{\rho_{1a+}}=\cos\left(  \theta_{b}-\theta_{a}\right)  /2$.

We see therefore that the value ascription given by Eq. (\ref{e33}) \emph{does
not fit} with the main property of the model \footnote{Since value ascriptions
given by $V(\Omega)$ or $p(V,\Omega)$ are characteristic of Bell models, it
could be said that even for a single particle, Bell-type models are
inconsistent with the present model. On the other hand, it could be argued
that the ensemble $\rho_{1}$ should be taken as the 'hidden variable', given
that this is the variable ascribing values to the outcomes and probabilities,
even though $\rho_{1}$ may not qualify following Bell's terminology as being a
'beable' (see Sect.\ V).}.\ The reason is that\ Eq. (\ref{50}) introduces an
ensemble dependency on the model: the probabilities do not depend on the
phase-space position but on the ensemble, as if the particle's angular
momentum effectively occupied an entire hemisphere (physically, this may
happen for example if the particle follows a stochastic motion with its
angular momentum constrained to remain in the ensemble, the timescale of the
measurement being significantly larger than the timescale of the stochastic
motion). Eq. (\ref{e33}) should thus be replaced by%
\begin{equation}
R_{1a}(\Omega_{1})=\pm1/2\Leftrightarrow J_{1a}\gtrless0\ \text{for
\emph{every} }J_{1a}\in\rho_{1}. \label{35}%
\end{equation}

\begin{figure}[tb]
\includegraphics[height=2.1in,width=5.2in]{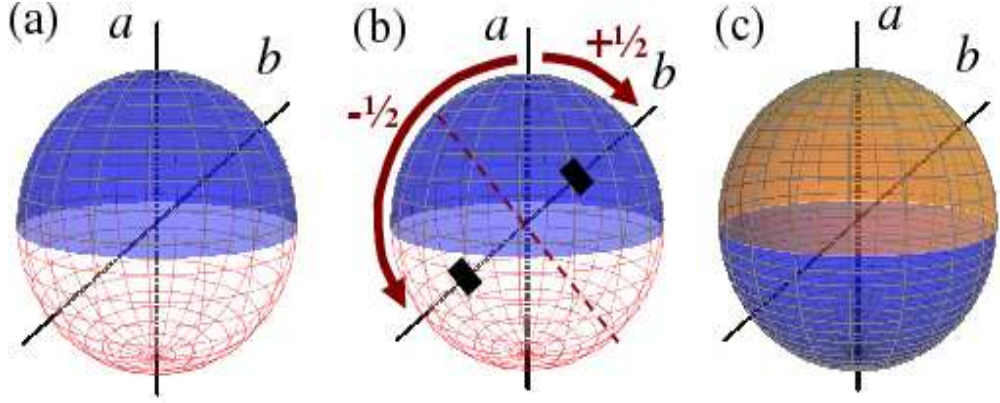}
%
%\label{f1}%
\caption{(a) The ensemble $\rho_{1a+}$ for the single particle model
described in Sec. IV.B. Any $\mathbf{J}_1$ in this ensemble has a
positive projection $J_{1a}$; measuring $R_{1a}$ gives the outcome
$+1/2$ with certainty, without changing the ensemble, since
$\rho_{1a+}$ is symmetric relative to the $a$ axis and $\left\langle
J_{1a}\right\rangle _{\rho_{1a+}}=1/2$. (b) In the same situation
$R_{1b}$ is measured. Now the symmetry axis of the ensemble
$\rho_{1a+}$ does not coincide with the $b$ axis. Hence measuring
$R_{1b}$ can yield either $+1/2$ or $-1/2$ with probabilities
proportional to $\left\langle H(\pm J_{1b})J_{1b}\right\rangle
_{\rho_{1a+}}$. The ensemble is modified during the measurement,
undergoing a rotation toward the positive or negative $b$ axis as
indicated by the arrows. (c) The two-particle distribution after
$R_{1a}$ was measured and the outcome is known to have been
$R_{1a}=-1/2$, in which case particle 1 is described by the ensemble
$\rho_{1a-}$. Conservation of the angular momentum then requires
that particle 2 be described by $\rho_{2a+}$, so that if $R_{2a}$
were measured, the outcome $+1/2$ would be obtained with certainty
[Eq. (\ref{33})]. If instead $R_{2b}$ is measured, we have a single
particle problem
for particle 2 identical to the one portrayed in Fig. 2(b). }%
\end{figure}

\subsubsection{Further considerations\label{mc}}

Although this has no effect on the computations, it will be
convenient, in order to provide a physical interpretation, to detail
the consequences arising from the model. Eq. (\ref{35}) associates
an outcome $R_{1a}$ along an axis $a$ with $J_{1a}$ being of the
same sign for every member of the hemispheric ensemble $\rho_{1a\pm
}$ (see Fig.\ 3(a)). Since $R_{1a}=\pm1/2=\left\langle
J_{1a}\right\rangle _{\rho_{1a\pm}}$, we can envisage that the
random interaction occurring during a measurement effectively
changes the distribution of the angular momentum: for instance if
initially the distribution is on a given hemisphere, say
$\rho_{1a+},$ Eq. (\ref{35}) is realized and $R_{1a}=1/2$ is
obtained with certainty, reflecting $\left\langle
J_{1a}\right\rangle _{\rho_{1a+}}.$ If $R_{1b}$ is measured, the
final distribution is $\rho_{1b+}$ (resp. $\rho_{1b-}$) if the
outcome $k=1/2$ (resp. $-1/2$) is obtained (see Fig.\ 3(b)). The
outcome thus appears as the average value of the angular momentum
projection in the post-measurement
distribution and Eq. (\ref{e25}) becomes%
\begin{equation}
\left\langle R_{1b}\right\rangle _{\rho_{1a+}}=\sum_{k=\pm1}\left\langle
J_{1b}\right\rangle _{\rho_{1b(k)}}P(R_{1b}=k,\rho_{1a+})=\left\langle
J_{1b}\right\rangle _{\rho_{1a+}}.\label{37}%
\end{equation}
Note that this implies that consecutive measurements involving projections
along different axes \emph{do not commute}: the condition (\ref{35}) cannot be
realized jointly along two different directions, like the classical analogues
of the angular momenta eigenstates presented in Sec.\ II \footnote{We have
chosen distributions on hemispheres rather than the ring distributions of
Sec.\ II for continuity with the examples investigated in Sec.\ III; the model
studied here would also hold if ring like distributions were employed.}. If
the initial distribution is $\rho_{1a+}$ measuring $R_{1b}$ then
$R_{1a^{\prime}}$ entails that $R_{1b}$ is measured over $\rho_{1a+}$ but
$R_{1a^{\prime}}$ over $\rho_{1b\pm}$ depending on the outcome $R_{1b}$. In
the reverse order, $R_{1a^{\prime}}$ is measured first, the average being
given by $\left\langle J_{1c}\right\rangle _{\rho_{1a+}}$ and $R_{1b}$ then
involves the values of $J_{1b}$ over one of the distributions $\rho
_{1a^{\prime}\pm}$. Eq. (\ref{37}) also allows to compute the change in the
angular momentum projection due the measurement,%
\begin{equation}
\Delta\left\langle J_{1b}\right\rangle \equiv\left\langle J_{1b}\right\rangle
_{\rho_{1b(2k)}}-\left\langle J_{1b}\right\rangle _{\rho_{1a+}}=-2kP(R_{1b}%
=k,\rho_{1a+}),\text{ }k=\pm\frac{1}{2}.\label{33m}%
\end{equation}

Consider now the uniform distribution on the entire sphere $\rho_{1\Sigma}$.
It can first be envisaged as the angular momentum occupying the upper or lower
hemispheres along a definite direction (say $a$) so that%
\begin{equation}
\rho_{1\Sigma}=(\rho_{1a+}+\rho_{1a-})/2.\label{32}%
\end{equation}
Since distributions in classical mechanics obey the principle of linear
superposition, $\rho_{1\Sigma}$ can also be taken as a sum of the expressions
given by Eq. (\ref{32}) over different directions $a$. Alternatively the angle
$a$ in Eq. (\ref{32}) can be taken to vary in time (then the measurement does
not involve a change in the distribution but rather a selection of the angular
momenta such that $J_{1a}>0$ or $J_{1a}<0$), or $\mathbf{J}$ be distributed on
the entire spherical surface (then the measurement induces a change in the
distribution $\rho_{1\Sigma}\rightarrow\rho_{1a\pm}$). Only in these latter
cases is the distribution spherically symmetric; all these possibilities lead
to the same probabilities and average values, yielding $P(R_{1a}=\pm
1/2,\rho_{1_{\Sigma}})=1/2$ for \emph{any} axis $a$ as well as a vanishing
average (\ref{50}) as required.

\subsection{Two-particle expectation}

\subsubsection{Distribution and conservation of the angular momentum}

Before computing the two-particle averages and correlation
functions, we explicitate the initial distribution and the
conservation of the angular momentum for the model. We have seen
that the defining property Eq. (\ref{50}) implied that value
ascription depended on distributions (taken to be ensembles on given
hemispheres) and not on individual phase-space positions. The
two-particle distribution given above by Eq. (\ref{9}) is (i)
spherically symmetric and (ii) anti-correlates the individual
positions of the angular-momenta $\mathbf{J}_{2}=-\mathbf{J}_{1}$,
so that we have $J_{2a}=-J_{1a}$ for projections along arbitrary
axes $a$ on the sphere. We require the extension of these two
properties so that they hold over the initial distribution, to be
denoted by $\rho_{\Sigma}.$ We must thus have for any
of the two particles $i$ and axis direction $a$%
\begin{equation}
\left\langle R_{ia}\right\rangle _{\rho_{\Sigma}}=\left\langle J_{ia}%
\right\rangle _{\rho_{\Sigma}}=0,\label{32a}%
\end{equation}
so that both outcomes $R_{ia}=\pm1/2$ can be obtained with equal probability.
The correlation between the outcomes for the two particles is obtained by
applying Eq. (\ref{50b}) to $\rho_{\Sigma,}$ giving%
\begin{equation}
\left\langle H(J_{2a})J_{2a}\right\rangle _{\rho_{\Sigma}}=\left\langle
H(-J_{1a})J_{1a}\right\rangle _{\rho_{\Sigma}}.\label{33e}%
\end{equation}
By Eq. (\ref{37}) we have $R_{1a}=\left\langle J_{1a}\right\rangle
_{\rho_{1\pm a}}$ so that by way of Eq. (\ref{33e}) the anti-correlation
$J_{2a}=-J_{1a}$ implies that the outcomes and the distributions for the
particles along the same axis are anti-correlated,%
\begin{equation}
\left\langle J_{2a}\right\rangle _{\rho_{2a\mp}}\equiv R_{2a}=-R_{1a}%
\equiv-\left\langle J_{1a}\right\rangle _{\rho_{1a\pm}}.\label{33}%
\end{equation}
Eqs. (\ref{32a})-(\ref{33}) hold for any arbitrary axis $a$. The
anti-correlation for the measurement outcomes, based on the
conservation of the angular momentum over the ensembles, implies
anti-correlations \emph{between these ensembles}. Measuring $R_{1a}$
links the outcome to one of the two ensembles $\rho_{1a\pm}$
depending on whether $R_{1a}=\pm1/2$. In turn, this also fixes
$\rho_{2}=\rho_{2a\mp}$. Note that contrarily to the correlation
between individual phase-space positions (for which one has
$J_{2a}=-J_{1a}$ and $J_{2b}=-J_{1b}$ for any axes $a$ and $b$), Eq.
(\ref{33}) cannot hold jointly along several directions.\ This is a
consequence of Eq. (\ref{35}) not holding simultaneously along
several axes.

There are different possibilities for choosing explicit realizations of
$\rho_{\Sigma}:$ all these possibilities lead to the same results and all
hinge on the conservation of the \emph{total }angular momentum along an
arbitrary axis demanded by Eq. (\ref{33}). For example $\rho_{\Sigma}$ can be
taken as proportional to $\rho_{1b+}\rho_{2b-}+$ $\rho_{1b-}\rho_{2b+}.$ Eq.
(\ref{33}) is then ensured provided the change in the angular momentum
(\ref{33m}) after the first measurement is taken into account in the angular
momentum balance for the other particle. Alternatively as in Eq. (\ref{32}),
$b$ can be taken as varying in time, giving
\begin{equation}
\rho_{\Sigma}=\frac{1}{2}\left(  \rho_{1b(t)+}\rho_{2b(t)-}+\rho_{1b(t)-}%
\rho_{2b(t)+}\right)  . \label{33c}%
\end{equation}
As for the single particle distribution $\rho_{1\Sigma}$, measuring $R_{1a}$
then selects the individual positions of $J_{1a}$ such that $J_{1a}\gtrless0$,
correlated to the individual positions $J_{2a}\lessgtr0$. Another possibility
for $\rho_{\Sigma}$ would be to take the distribution (\ref{9}) and consider
$R_{1a}$ as inducing a change in the distribution $\rho_{\Sigma}%
\rightarrow\rho_{1a\pm}$.

\subsubsection{Computation of the correlation}

Since the measurement outcomes do not depend on the individual phase-space
positions, the average $E(a,b)\equiv\left\langle R_{1a}R_{2b}\right\rangle
_{\rho_{\Sigma}}$ cannot be obtained as in the preceding example from the
phase-space averages (\ref{e17})-(\ref{e20}), but from the probabilities of
detecting a given outcome as a function of the distribution. $E(a,b)$ is
computed from the general formula, also employed in Sec. III.C%
\begin{equation}
\left\langle R_{1a}R_{2b}\right\rangle _{\rho_{\Sigma}}=\sum_{k,k^{\prime
}=-1/2}^{1/2}kk^{\prime}P_{kk^{\prime}} \label{e38}%
\end{equation}
where as in Eq. (\ref{e21}) $P_{kk^{\prime}}$ is given by%
\begin{equation}
P_{kk^{\prime}}=P(R_{1a}=k\cap R_{2b}=k^{\prime},\rho_{\Sigma})=P(R_{1a}%
=k)P(R_{2b}=k^{\prime}|R_{1a}=k) \label{e39}%
\end{equation}
and the two particle expectation takes the form%
\begin{equation}
\left\langle R_{1a}R_{2b}\right\rangle _{\rho_{\Sigma}}=\sum_{k=-1/2}%
^{1/2}kP(R_{1a}=k)\left[  \sum_{k^{\prime}=-1/2}^{1/2}k^{\prime}%
P(R_{2b}=k^{\prime}|R_{1a}=k)\right]  . \label{e40}%
\end{equation}
For any particle $i$ and direction $a$, we have
\begin{equation}
P(R_{ia}=\pm1/2,\rho_{\Sigma})=1/2. \label{e31}%
\end{equation}
The conditional probability $P(R_{2b}=k^{\prime}|R_{1a}=k)$ is, as in the
example involving discrete outcomes studied above in Sec.\ \ref{exd}, the
probability of obtaining $R_{2b}=k^{\prime}$ if it known that $R_{1a}=k$. But
we have just seen that obtaining an outcome $R_{1a}=k$ is linked to the
respective densities $\rho_{1a[\mathrm{sign}(k)]}$ and $\rho_{2a[\mathrm{sign}%
(-k)]}.$ The conditional probability is therefore given by%
\begin{equation}
P(R_{2b}=k^{\prime}|R_{1a}=k)=P(R_{2b}=k^{\prime},\rho_{2a[\mathrm{sign}%
(-k)]}), \label{31b}%
\end{equation}
which is a single particle probability of the type\ given by Eq.
(\ref{esp}). Note that in \ order to compute the expectation value,
we do not need to know the values of these individual probabilities,
as the knowledge of the conditional expectation -- the expression
between brackets in Eq. (\ref{e40}) -- is sufficient. The 2-particle
conditional expectation is given by the single particle average
$\left\langle J_{2b}\right\rangle _{\rho _{2a[\mathrm{sign}(-k)]}}$
whose expression was determined above (Eqs. (\ref{50}), (\ref{e25})
and (\ref{e30})). We can rewrite the average in the
form%
\begin{equation}
\sum_{k^{\prime}=-1/2}^{1/2}k^{\prime}P(R_{2b}=k^{\prime}|R_{1a}%
=k)=-k\cos(\theta_{b}-\theta_{a}). \label{e44}%
\end{equation}

We can now compute the expectation $E(a,b)\equiv\left\langle R_{1a}%
R_{2b}\right\rangle $ from Eqs. (\ref{e40}) and (\ref{e44}). The result is
easily seen to be
\begin{equation}
E(a,b)=-\frac{1}{4}\cos(\theta_{b}-\theta_{a}). \label{55}%
\end{equation}
In the present derivation, we have assumed that the knowledge of particle 1's
outcome was obtained first, hence the appearance of the conditional
probability regarding the outcomes of particle 2. But obviously by Bayes'
theorem the result is the same if we assume instead that $R_{2b}$ is known
first, and the conditional probability then concerns the computation of the
outcomes of particle 1.

The correlation function $C(a,b,a^{\prime},b^{\prime})$ is again given by Eq.
(\ref{23}) with $V_{\max}=1/2$.\ $C(a,b,a^{\prime},b^{\prime})$ violates the
Bell inequality (\ref{25}) for a wide range of values, the maximal violation
being obtained for $C(0,\frac{\pi}{4},\frac{\pi}{2},\frac{3\pi}{4})=2\sqrt{2}%
$. This correlation function, with $E(a,b)$ given by Eq. (\ref{55}), is
familiar from quantum mechanics -- it is precisely the correlation obtained
for the 2 particles with spin $1/2$ in the singlet state. It was shown in this
case that Eq. (\ref{55}) can be seen as a consequence of a particular
correlation between vectors whose projection is conserved on average
\cite{unnik05}.

\section{Discussion}

\subsection{Ensemble dependence}

We have seen in our fourth example (Sec. IV) that correlation functions
obtained from 2-particle distributions in classical mechanics can lead to a
violation of the Bell inequalities, without nonlocality being explicitly
involved (it may play an implicit role, see Sec.\ C below). The main
difference between this model and the other examples we have given consists in
the ensemble dependencies: probabilities, average values and conservation laws
are relative to a collective property (a given distribution) and do not
depend, as in the other cases, on the individual phase-space positions.
Indeed, the constraint (\ref{50}) cuts the link between a definite phase-space
position of a particle and a given measurement outcome (be it in a
probabilistic or deterministic way).

In this respect, it is noteworthy to compare the interpretation of the
conditional probabilities appearing in Examples 2 (Sec.\ III.C) and 4
(Sec.\ IV). In both cases $P(V_{2b}=k^{\prime}|V_{1a}=k)$ is grounded on the
correlation (\ref{e1}) and represents the probability of obtaining
$V_{2b}=k^{\prime}$ given the knowledge that $V_{1a}=k$. In both cases the
distribution of $\mathbf{J}_{2}$ is modified once the outcome $V_{1a}=k$ is
known \footnote{It seems it is necessary to stress that the change in the
probability distribution of particle 2 when the outcome of particle 1 is known
is not a physical phenomenon involving action at a distance, but the result of
the information brought by the knowledge of the first outcome, given the
conservation law. This point, unrelated to Bell's theorem, is an elementary
inference common in the calculus of probabilities.} (it changes from a uniform
distribution on the sphere to a uniform distribution on the positive or
negative hemisphere centered on $a$, depending on $k$). However in example 2
the probabilities depend on the individual phase-space positions of the
particles: although it may be unknown in practice, $\mathbf{J}_{1}$ has in
principle a definite position that determines $V_{1a}=k$, and to this position
corresponds the definite position $\mathbf{J}_{2}=-\mathbf{J}_{1}$ that will
determine the outcome $V_{2b}$; so the conditional probability is computed by
finding the individual positions of $\mathbf{J}_{2}$ such that $V_{2b}%
=k^{\prime}$ compatible with the positions of $\mathbf{J}_{1}$ imposed by
$V_{1a}=k$ (namely $J_{1a}>0$). In example 4 an outcome $V_{1a}=k$ cannot be
linked in principle to an individual position of $\mathbf{J}_{1}$ and thus we
can only infer from the outcome the \emph{ensemble} to which $\mathbf{J}_{1}$
must belong; then from the conservation law we know the distribution for
$\mathbf{J}_{2}$, which allows to compute $P(V_{2b}=k^{\prime}|V_{1a}=k)$ from
the probability $P(R_{2b}=k^{\prime},\rho_{2a[\mathrm{sign}(-k)]}).$ Hence we
can only correlate observable outcomes with ensembles, not with individual
positions of the angular momenta. Assuming that a given phase-space position
determines probabilities, as in the stochastic model of Sec. III.D, only
brings in several combinations of possible outcomes allowed by the definite
positions of $\mathbf{J}_{1}$ and $\mathbf{J}_{2}=-\mathbf{J}_{1}$ on the
angular momentum sphere, but still allows to correlate these individual
positions with measurement outcomes.

\subsection{Joint distributions and non-commutative measurements}

We had remarked in Sec.\ III.B that the existence of a joint
probability $P_{aba^{\prime}b^{\prime}}$ is sufficient to ensure
that a Bell-type inequality holds, irrespective of whether the
assumption that measurement outcomes and probabilities depend on the
individual phase space positions is made. But if that specific
assumption is made, then one is lead to the
existence of a joint probability by imposing the factorization (\ref{w1}%
).\ Along these lines, there are two ways of seeing why Bell's theorem does
not apply to our fourth example.

First, the ensemble dependence can formally be thought of as arising
from elementary phase-space probability functions \emph{specific} to
a given ensemble,
i.e.%
\begin{equation}
P(R_{1a}=k,\rho_{1})=\int p(\Omega_{1};\rho_{1})\rho_{1}(\Omega_{1}%
)d\Omega_{1} \label{e56}%
\end{equation}
(compare with Eq. (\ref{w6})). By employing expressions such as Eq.
(\ref{e56}) in the expectation value as given by Eqs. (\ref{e38})-(\ref{e39}),
it can be seen directly that the ensemble dependence of the elementary
probabilities spoils the factorization (\ref{w1}) -- for example instead of
$p(\Omega_{2}),$ one has outcome dependent expressions such as $p(\Omega
_{2};\rho_{2}(R_{1a}))$.

The second manner starts with the remark made above concerning the
non-commutation of the $R$ measurements introduced in our model; in
classical mechanics, measurements usually commute, but this is not
the case if they arise from collective phenomena (encapsulated in
the ensemble dependency). By requiring that the angular momentum be
conserved between ensembles (just as it is when individual positions
are considered), the consequences of the non-commutation are carried
over from one particle to the other. As seen in Sec. \ref{mc} in the
single particle case, the probabilities and outcomes when $R_{1b}$
is measured after a first measurement is made will be different
depending on whether $R_{1a}$ or $R_{1a^{\prime}}$ was measured
first. Because Eq. (\ref{33}) links the outcomes with the ensembles,
this is also the case in the two-particle problem when $R_{2b}$ (or
$R_{2b^{\prime}}$) is determined after $R_{1a}$ \emph{or}
$R_{1a^{\prime}}$ were measured. Put differently, although Eq.
(\ref{33}) holds for the $a$, $a^{\prime}$, $b$ and $b^{\prime}$
axes, it cannot hold jointly for all the axes because the single
particle ensembles $\rho_{1a\pm}$ and $\rho_{1a^{\prime}\pm}$ are
mutually exclusive, as well as $\rho_{2b\pm}$ and
$\rho_{2b^{\prime}\pm}$ (see Sec. II and Sec. IV.A.2). Hence a joint
probability $P_{aba^{\prime}b^{\prime}}$ cannot be defined and the
model is not constrained by the inequality (\ref{i1}). This is
consistent with the equivalence \cite{fine82,malley04} shown between
the verification of Bell's inequality and the commutation of the
four observables entering Eq. (\ref{e60}). The ensemble-dependence
introduced in our model appears as a tool in order to enforce, in a
classical context, the non-commutation of the measurements along
different axes made on the same particle.

\subsection{Conservation laws and locality}

Factorization, enforcing the existence of joint distributions, and
as such a necessary assumption in the derivation of Bell's theorem,
is usually argued to be intimately linked to locality. According to
Bell \cite{bellNC}, factorization is a \emph{consequence} of local
causality, given that space-like separated events can only have
common causes in their backward light-cone: therefore the
probability of obtaining a certain result in an event regarding one
of the particles cannot depend on what has been measured on the
other. It is known however that factorization can be seen
\cite{jarrett,shimony} as a consequence of two separate conditions,
outcome independence (the conditional probability of one event does
not depend on the outcome obtained in the other event) and parameter
independence (dependence on the measurement direction of the other
event). Only the violation of outcome independence can result in a
genuine violation of local causality (it would permit superluminal
signalling), whereas the violation of parameter independence allows
a 'peaceful coexistence' \cite{shimony} between local causality and
other types of\ correlations preventing the factorization.

The present model -- like many quantum mechanical entangled systems
-- respects parameter independence [Eq. (\ref{e31})] but violates
outcome independence [Eq. (\ref{31b})] (the dependencies here must
be understood relative to the ensembles and not relative to the
individual positions of the angular momenta). This outcome
dependence of the conditional probabilities is due to the
conservation of the angular momentum, as encapsulated by
$\mathbf{J}_{2}=-\mathbf{J}_{1}$ (anti-correlation between
individual positions), Eq. (\ref{33e}) (correlation between
ensembles occupying opposite hemispheres centered on the same
arbitrary axis) and Eq. (\ref{33}) (anti-correlation between the
outcomes made on the same arbitrary axis). Parameter independence on
the other hand guarantees that the predictions relative to $R_{1a}$
do not depend on what measurement or whether a measurement is
carried out on particle 2, and vice-versa [Eq$.$ \ref{e31})]. It is
clear nevertheless that the angular momentum conservation affects
the distributions of both particles. For example if $\rho_{\Sigma}$
is given by Eq. (\ref{33c}), made up from rotating distributions,
measuring $R_{1a}$ not only freezes the rotation of particle's 1
distribution, but also that of particle 2 (precisely because the
angular momenta are correlated and need to be conserved). If
$\rho_{\Sigma}$ is taken as a uniform distribution on the sphere for
the individually anti-correlated angular momenta, measuring $R_{1a}$
changes the distribution $\rho_{1\Sigma}\rightarrow\rho_{1a\pm}$ but
also $\rho_{2\Sigma}\rightarrow\rho_{2a\mp}$. Hence, it can be
argued that the conservation of the angular momentum as implemented
in our model actually results from an implicit implementation of
nonlocality. There are several answers to this question, depending
on the status one gives to conservations laws and ensemble
distributions, or on how nonlocality or causality are defined. The
three following positions can be singled out:

\begin{enumerate}
\item The changes in the distributions are real physical effects, but the
conservation of the angular momentum results from a symmetry that is
intrinsically linked to space-time. Indeed the correlation
(\ref{33}) arises by generalizing the angular momentum conservation
for individual positions to ensembles accounting for non-commutative
measurements. There is no need to invoke a specific mechanism for a
conservation law -- conservation laws and symmetry principles are
just postulated. But if desired, a field can be can be ascribed the
role of transporting the angular momentum; in this respect, it may
be useful to make the analogy with Feynman's paradox in which
mechanical angular momentum is transmitted between two charged
particles through the electromagnetic field \cite{fp}.

\item The changes in the distributions are real physical effects due to a
nonlocal form of causation. The requirement given by Eq. (\ref{33})
is sufficient to imply nonlocality. Action at a distance effects are
quite common in non-relativistic classical mechanics, although the
modern view is to ascribe such effects (like gravity or several
phenomena in electrostatics) to the action of fields. Here the
nonlocal effect would consist in accounting for angular momentum
conservation.\ This does not necessarily contradict the preceding
position since it can be argued that symmetries can give rise to
nonlocality, a position leading to a holistic vision of symmetries
as holding beyond a space-time framework.

\item The changes in the distributions are not physical effects: one must
distinguish the observed frequencies (which are measured) from the
calculus of probabilities (whose role is to make logical inferences
given a certain state of information \cite{jaynes}). Conditional
probabilities do not therefore express causation and the
factorization of the probabilities does not follow from the
requirement of local causality.\ In Bell's term, the variables
entering the probabilities are not beables, an argument that may be
supported by the fact that individual angular momenta positions do
not ascribe values and that the status of ensembles as beables is
questionable. The ensembles and their correlations are theoretical
constructs encapsulating the state of knowledge we have of the
situation, including the constraints (like conservation laws).
\end{enumerate}

\section{Conclusion}

To summarize, we have first constructed classical distributions
analogues of the quantum mechanical angular momentum eigenstates;
these classical distributions are characterized by being mutually
exclusive, leading, with appropriate assumptions to non-commutative
measurements. We have then derived Bell's theorem in the
deterministic and stochastic cases; both cases are characterized by
the fact that an individual position of the angular momentum
ascribes a value (with certainty or with a given probability) to a
measurement of the projection along any axis. As a result, a joint
probability distribution for an arbitrary number of events can be
defined. Three different examples of Bell-type models were studied.
A non-Bell-type model was introduced in Sec.\ IV: in this model,
individual positions of the angular momenta are irrelevant to
determine the measurement outcomes, that only depend on ensembles.\
As a result, single particle measurements do not commute, and a
distribution for joint measurements along different axes cannot be
defined. If it further assumed that the total angular momentum must
be conserved, the Bell inequalities are violated.

The present results do not disprove Bell's theorem -- as we have seen, in
these circumstances the assumptions made in the derivation of the theorem are
not fulfilled. We have argued that the violation of the inequalities in our
classical model is due to the conservation of the total angular momentum in
the context of non-commutative measurements; nonlocality does not need to be
invoked (although it may).\ From this perspective, the violation of the Bell
inequalities would not constitute a marker of nonlocality. It still remains to
be investigated what type of collective or individual phenomena are compatible
with the type of model introduced in this work.

\end{document}